\documentclass[aps,prd,twocolumn,superscriptaddress,byrevtex,longbibliography]{revtex4-2}
\usepackage[utf8]{inputenc}
\usepackage{color}
\usepackage{graphicx}

\usepackage{amsmath,amssymb,amsfonts}

\def\prl{Phys. Rev. Lett.}
\def\prd{Phys. Rev. D}

\def\apjl{Astrophys. J. Lett.}
\def\aap{Astronomy and Astrophysics}

\def\mnras{Mon. Not. R. Astron. Soc.}

\def\physscr{Physica Scripta}
\def\jcap{JCAP}

\newcommand\longcomment[1]{}

\begin{document}

\title{Neutron Stars Harboring a Primordial Black Hole: Maximum Survival Time}

\author{Thomas W.~Baumgarte}

\affiliation{Department of Physics and Astronomy, Bowdoin College, Brunswick, Maine 04011}

\author{Stuart L.~Shapiro}

\affiliation{Department of Physics, University of Illinois at Urbana-Champaign, Urbana, Illinois 61801}

\affiliation{Department of Astronomy and NCSA, University of Illinois at Urbana-Champaign, Urbana, Illinois 61801}

\begin{abstract}
We explore in general relativity the survival time of neutron stars that host an endoparasitic, possibly primordial, black hole at their center.  Corresponding to the minimum steady-state Bondi accretion rate for adiabatic flow that we found earlier for stiff nuclear equations of state (EOSs), we derive analytically the maximum survival time after which the entire star will be consumed by the black hole. We also show that this maximum survival time depends only weakly on the stiffness for polytropic EOSs with $\Gamma \geq 5/3$, so that this survival time assumes a nearly universal value that depends on the initial black hole mass alone. Establishing such a value is important for constraining the contribution of primordial black holes in the mass range $10^{-16} M_\odot \lesssim M \lesssim 10^{-10} M_\odot$ to the dark-matter content of the Universe.
\end{abstract}

\maketitle

Primordial black holes (PBHs) that may have formed in the early Universe (see, e.g., \cite{Haw71,CarH74}) have long been considered candidates for contributing to, if not accounting for, the mysterious and elusive dark matter (see, e.g., \cite{Cha75}, as well as \cite{CarK20} for a recent review).  Constraints on the PBH contribution to the dark matter have been established by a number of different observations.  Sufficiently small PBHs ($M \lesssim 5 \times 10^{14}$ g) would have evaporated due to Hawking radiation \cite{Haw74} within less than a Hubble time, while, for larger black-hole masses, different types of observations have resulted in limits for different mass ranges (see, e.g., \cite{CapPT13,KueF17} and references therein; see also \cite{SasSTY18} for a review of constraints arising from gravitational-wave observations, as well as \cite{VasV21} for a possible detection by NANOGrav).

One compelling constraint on the PBH contribution to the dark matter in the mass range $10^{-16} M_\odot \lesssim M \lesssim 10^{-10} M_\odot$ (which currently is not well-constrained by other observations; see, e.g., \cite{CarKSY20,CarK20}) results from the fact that PBHs can be captured by stars, and would then accrete and swallow these stars (see, e.g., \cite{Haw71,Mar95}).  This process may be particularly efficient for capture by neutron stars (see, e.g., \cite{CapPT13,GenST20}, but see also \cite{MonFVSH19}), so that the existence of neutron-star populations provides a limit on the density of PBHs, and hence on their contribution to the dark-matter content of the Universe.  Other observational signatures of this process have been discussed, for example, in \cite{FulKT17,TakFK20}.

Evidently, the above argument can provide constraints on PBHs only if the capture and subsequent accretion-driven destruction of the neutron star proceeds on time\-scales shorter than the age of the oldest neutron-star populations; it therefore hinges on reliable time-scale estimates for these processes.  Estimates for the capture of PBHs by neutron stars via gravitational focusing, followed by the dissipation of PBH kinetic energy via dynamical friction, accretion, surface and gravitational waves that lead to the settling down of the PBH near the center of the neutron star, have recently been revisited by \cite{GenST20,MonFVSH19}.  The subsequent accretion rate of the star by the endoparasitic black hole is often
estimated analytically (see \cite{EasL19,RicBS21b} for numerical simulations) by adopting the spherical, steady-state, Bondi accretion formula,
\begin{equation} \label{mdot_bondi}
    \dot M = 4 \pi \lambda \frac{M^2 \rho_0}{a^3}
\end{equation}
for adiabatic flow (\cite{Bon52}; see also \cite{ShaT83} for a textbook treatment, including its relativistic generalization). Here $\rho_0$ and $a$ are the ``asymptotic" rest-mass density and the sound speed, respectively, measured at large distances from the black hole, but, for small black holes, still well inside the neutron star's nearly homogeneous central core. The mass of the gas within the accretion radius is assumed to be negligible in comparison with the black hole mass. The parameter $\lambda$ is a dimensionless ``accretion eigenvalue", which, for $1 \leq \Gamma \leq 5/3$, is a constant of order unity.

In a general relativistic formulation, and for accretion onto a black hole at the center of a neutron star, the dot in (\ref{mdot_bondi}) denotes a derivative with respect to time as measured by a ``local asymptotic" static observer far from the black hole but again well inside the neutron star core (see Appendix A in \cite{RicBS21b} for a full derivation of the spacetime metric and applicability of the relativistic Bondi flow solution; see footnote \footnote{The underlying reasons for this result are that (1) the self-gravity of the neutron star gas that is bound to the black hole inside $r_a \sim M/a^2 \ll R$ can be neglected in comparison to the gravitational field of the black hole,  and (2) the (spherical) stellar matter exterior to $\sim r_a$ has no influence on the accretion flow by virtue of Birkhoff's theorem.}).  This time advances at a rate somewhat slower than that measured at infinity due to the gravitational redshift from the  core.  This redshift factor is given by the ``lapse" function $\alpha$ for the local asymptotic observer, typically $0.6 \lesssim \alpha < 1$.  The accreted mass measured by (\ref{mdot_bondi}) is fundamentally a (baryon) {\it rest} mass $dM_{\rm rest}$, which enhances the black hole's gravitational mass $M$ by a somewhat larger amount due to the accretion of additional internal energy, $\beta \equiv dM / dM_{\rm rest} \gtrsim 1$  (see \cite{RicBS21b}).  Accounting for both of these factors,  we may rewrite the accretion rate (\ref{mdot_bondi}) as
\begin{equation} \label{mdot_bondi_rel}
    \frac{d M}{dt} = 4 \pi \alpha \beta \lambda \frac{M^2 \rho_0}{a^3},
\end{equation}
where $t$ now measures the time as observed by an observer at infinity, and $M$ is the black hole's gravitational mass.  Since the product of the two terms $\alpha \lesssim 1$ and $\beta \gtrsim 1$ results in a dimensionless number very close to unity, $\alpha \beta \approx 1$, we will ignore this product in the following.

For accretion onto black holes, a relativistic treatment for Schwarzschild black holes shows that the requirement that the sound speed be less than the speed of light demands that the flow pass through a critical point, yielding a unique value for $\lambda$ \cite{ShaT83}. Adopting a polytropic equation of state (EOS)
\begin{equation} \label{polytrope}
    P = K \rho_0^{\Gamma},
\end{equation}
where $K$ is a constant, $\Gamma = 1 + 1/n$ is the adiabatic index and $n$ is the polytropic index, that unique value for $1 \leq \Gamma \leq 5/3$, assuming $a \ll 1$, is
\begin{equation} \label{lambda}
    \lambda = \frac{1}{4} \left( \frac{2}{5 - 3 \Gamma} \right)^{(5 - 3 \Gamma)/2 (\Gamma - 1)},
\end{equation}
which is the value found in the Newtonian formulation.  Note that here and throughout we adopt geometrized units with $G~=~1~=~c$.

While the accretion timescale may be small compared to the total capture and settling time for many black hole masses, there is some uncertainty in determining the actual accretion timescale when applying the Bondi accretion rate given by Eq.~(\ref{mdot_bondi_rel}).  We first observe that (\ref{mdot_bondi_rel}) depends on both $\rho$ and $a$, which may vary from star to star (even though their order of magnitude is probably similar for all neutron stars).  More importantly, the accretion eigenvalues $\lambda$ are easily derived only for soft EOSs with $1 \leq \Gamma \leq 5/3$, as in Eq.~(\ref{lambda}).  As a result, some authors (e.g. \cite{KouT14,GenST20}) have resorted to approximating the accretion rate by adopting values for $\Gamma$ such as $\Gamma = 4/3$, even though one expects that the neutron star interior is governed by a significantly stiffer EOS with $\Gamma \gtrsim 2$, for which (\ref{lambda}) clearly does not apply. Finally, to make matters worse, consider that the polytropic EOS~(\ref{polytrope}) implies that the sound speed is given by
\begin{equation} \label{sound}
    a^2 \simeq \Gamma K \rho_0^{\Gamma - 1}
\end{equation}
whenever $P \ll \rho$, where $\rho$ is the total mass-energy density. This is the standard Newtonian relation. Inserting (\ref{sound}) into (\ref{mdot_bondi_rel}) results in 
\begin{equation} \label{mdot_a}
    \frac{dM}{dt} \simeq 4 \pi \lambda M^2 a^{(5 - 3 \Gamma)/(\Gamma - 1)}.
\end{equation}
We now observe that, for $a \rightarrow 0$, (\ref{mdot_a}) suggests that $dM/dt$ becomes infinite whenever $\Gamma > 5/3$, seemingly making this expression totally unreliable for stiff EOSs.

In this short paper we clarify this issue.  Building on our relativistic treatment of Bondi accretion for stiff polytropic EOSs (see \cite{RicBS21a}) we use our finding there that there exists a finite, nonzero, {\em minimum accretion rate} whenever $\Gamma \geq 5/3$ to show that, under quite general conditions, there exists a corresponding {\em maximum accretion time} for a PBH to swallow a neutron star. This maximum accretion time depends only weakly on the stiffness for EOSs with $\Gamma \geq 5/3$, thereby providing a nearly {\em universal} estimate, depending only on the initial black-hole mass, for the maximum time that a neutron star can survive accretion by an endoparasitic black hole.

We first observe that, for $\Gamma = 5/3$, the accretion rate (\ref{mdot_a}) becomes {\em independent} of the sound speed $a$.  A relativistic treatment shows that this value represents a {\em minimum accretion rate}, since relativistic corrections for higher sound speeds $a \lesssim 1$ and correspondingly larger densities lead to higher accretion rates.  As shown in \cite{RicBS21a}, Bondi accretion in general relativity exhibits such a minimum accretion rate for stiffer EOSs as well. This minimum rate is given by 
\begin{equation} \label{mdot_min}
    \left( \frac{dM}{dt} \right)_{\rm min} = \frac{4 \pi \bar \lambda}{\Gamma^{1/(\Gamma - 1)}} \, \frac{M^2}{K^{1/(\Gamma - 1)}},
\end{equation}
for all $5/3 \leq \Gamma \leq 3$.  Here $\bar \lambda$ is defined as
\begin{equation} \label{lambda_bar}
\bar \lambda = \bar x_{\rm III}^{(5 - 3\Gamma)/(2 \Gamma - 2)}
    \left( \frac{\Gamma - 1}{\Gamma - 1 - \bar x_{\rm III}}\right)^{1/(\Gamma - 1)} 
    \frac{(1 + 3 \bar x_{\rm III})^{3/2}}{4},
\end{equation}
with
\begin{equation}
\bar x_{\rm III} = \Gamma - 7/6 - \left(12 \Gamma - 11 \right)^{1/2} / 6,
\end{equation}
where we have adopted the notation of \cite{RicBS21a}. The existence of the minimum accretion rate (\ref{mdot_min}) results from the fact that, for $\Gamma > 5/3$ and to leading order in $a$, the accretion eigenvalues $\lambda$ can be written as $\lambda = \bar \lambda \, a^{(3 \Gamma - 5)/(\Gamma - 1)}$, so that the dependence on $a$ in (\ref{mdot_a}) cancels out, thereby leaving $\dot M$ finite but non-zero.  Note that (\ref{mdot_min}) reduces to (\ref{mdot_a}) for $\Gamma = 5/3$, in which case $\lambda = \bar \lambda = 1/4$.

Since the black hole accretes matter at a rate greater than the minimum rate (\ref{mdot_min}),
we can find the maximum time by integrating \footnote{For sufficiently small initial black holes, $M / M_{\rm tot} \ll 1$, the longest time is spent undergoing quasistatic Bondi accretion, while the final phase, when $M / M_{\rm tot} \sim 1$, only lasts a much shorter dynamical timescale; see \cite{RicBS21b}.}
\begin{eqnarray} \label{tmax0}
    t_{\rm max} & = & \int_{M_0}^{M_{\rm tot}} \frac{dM}{dM /dt} \leq
    \int_{M_0}^{M_{\rm tot}} \frac{dM}{(d M / dt)_{\rm min}} \nonumber \\
    & = &  
    \kappa K^{1/(\Gamma - 1)} \int_{M_0}^{M_{\rm tot}} \frac{dM}{M^2} 
    \nonumber \\
    & = & \kappa K^{1/(\Gamma - 1)} \left(\frac{1}{M_0} - \frac{1}{M_{\rm tot}} \right).
\end{eqnarray}
Here $M_0$ is the initial black-hole mass, $M_{\rm tot}$ is the total gravitational mass of the system, and we have defined
\begin{equation} \label{kappa}
 \kappa =\frac{\Gamma^{1/(\Gamma - 1)}}{4 \pi \bar \lambda}.
\end{equation}
Assuming that $M_0 \ll M_{\rm tot}$ we may neglect the last term in (\ref{tmax0}) and obtain
\begin{equation} \label{tmax1}
    t_{\rm max} = \kappa \frac{K^{1/(\Gamma - 1)}}{M_0}.
\end{equation}
This maximum time is dominated by the initial phase of accretion, when $M$ and the accretion rate given by Eq.~(\ref{mdot_min}) assume their lowest values. During this phase the accretion proceeds in a quasistationary fashion and the neutron star maintains its initial state (e.g., $\Gamma$ and $K$) to good approximation.

\begin{table}[!t]
    \centering
    \begin{tabular}{c|c|c|c|c|c}
        $n$ &  
        $\Gamma$ &
        $\bar \lambda$ &
        $\kappa$ &
        $\bar M_{\rm max}$ &
        $\kappa / \bar M_{\rm max}^2$ \\
        \hline
        1.5 &        1.67 &        0.25 &        0.685 &        0.265 &         9.75 \\
        1.4 &        1.71 &        0.432 &         0.392 &        0.237 &        6.98 \\
        1.3 &        1.77 &        0.640 &        0.261 &        0.213 &        5.75 \\
        1.2 &        1.83 &        0.889 &        0.185 &        0.194 &        4.92 \\
        1.1 &        1.91 &        1.17 &        0.138 &        0.177 &        4.40 \\
        1.0 &        2.0 &        1.49 &        0.107 &        0.164 &        3.96 \\
        0.9 &        2.11 &        1.83 &        0.085 &        0.152 &        3.69 \\
        0.8 &        2.25 &        2.15 &        0.071 &        0.142 &        3.46 \\
        0.7 &        2.43 &        2.45 &        0.060 &        0.135 &        3.32 \\
        0.6 &        2.67 &        2.69 &        0.053 &        0.129 &        3.21 \\
        0.5 &        3.0 &        2.83 &        0.049 &        0.125 &        3.12 
    \end{tabular}
    \caption{Polytropic parameters.  Here $n$ is the polytropic index, $\Gamma = 1 + 1/n$ the adiabatic index, the coefficients $\bar \lambda$ and $\kappa$ are defined in Eqs.~(\ref{lambda_bar}) and (\ref{kappa}), and $\bar M_{\rm max}~=~K^{-2/n} M_{\rm max}$ is the maximum dimensionless gravitational mass allowed for non-rotating neutron stars.  Note that the combination $\kappa / \bar M_{\rm max}^2$, which appears in the maximum accretion time (\ref{t_max}), depends only weakly on $\Gamma$ for stiff EOSs (with $\Gamma \gtrsim 2$, say). }
    \label{tab:parameters}
\end{table}

We can now determine the value of the constant $K$ for any polytropic EOS by matching its  dimensionless maximum mass $\bar M_{\max} = K^{-1/(2\Gamma - 2)} M_{\rm max}$ to the best current value value for the maximum mass of a nonrotating, isolated neutron star, $M_{\rm max}$.  The quantity $\bar M_{\rm max}$ is obtained by setting $K$ equal to unity and integrating the Tolman-Oppenheimer-Volkoff equations (see \cite{OppV39,Tol39}) to find the mass at the turning point along an equilibrium sequence of stars parametrized by their central density. We can therefore express $K$ as
\begin{equation} \label{K}
K = \left( \frac{M_{\rm max}}{\bar M_{\rm max}} \right)^{2(\Gamma - 1)}.
\end{equation}
Inserting (\ref{K}) into (\ref{tmax1}) we now obtain
\begin{equation} \label{tmax2}
    t_{\rm max} = 5 \times 10^{-6} \mbox{s} \,\frac{\kappa}{\bar M_{\rm max}^2} 
    \left( \frac{M_{\rm max}}{M_{\odot}} \right)^{2}
    \left( \frac{M_\odot}{M_{0}} \right), 
\end{equation}
where we have used that, in geometrized units, $1 M_{\odot} \simeq 5 \times 10^{-6}$ s.

We list values for the polytropic parameters appearing in (\ref{tmax2}) in Table \ref{tab:parameters}. Note that, for moderately stiff EOSs with $n \lesssim 1.0$ and $\Gamma \gtrsim 2$, say, the factor $\kappa / \bar M_{\rm max}^2$ varies only very moderately, with values between 3 and 4.  Adopting the maximum of these values,
\begin{equation} \label{kappa_max}
    \left. \frac{\kappa}{\bar M_{\rm max}^2} \right|_{\rm max} 
    \simeq 4,
\end{equation}
we can rewrite (\ref{tmax1}) as
\begin{equation} \label{t_max}
    t_{\rm max} \simeq 2 \times 10^{5} \mbox{s} \left( \frac{M_{\rm max}}{M_{\odot}} \right)^{2}
    \left( \frac{10^{-10} M_\odot}{M_{0}} \right).
\end{equation}

We next observe that, according to Fig.~4 of \cite{RicBS21a}, accretion rates for $\Gamma > 5/3$ depend only very weakly on the asymptotic sound speed, meaning that the actual accretion rate is comparable to its minimum value, and hence the actual accretion time is comparable to its maximum value (\ref{t_max}).  

For concreteness, we shall take the maximum mass of a nonrotating neutron star to be $M_{\rm max} = 2.3 M_{\odot}$. This mass is consistent with the likely range of values estimated by several groups (see, e.g., \cite{MarM17,ShiFHKKST17,RezMW18,RuiST18}) using data from the binary neutron star merger event GW170817 detected by LIGO/Virgo in gravitational waves \cite{LIGO17a}, as well as from the counterpart gamma-ray burst GRB17081A and kilonova AT 2017gfo. It is also consistent with the measurements of \cite{Croetal20}, which revealed the highest pulsar mass to date.  We then may rewrite (\ref{t_max}) as
\begin{equation} \label{t_max4}
    t_{\rm max} \simeq 1 \times 10^6 \mbox{s} \left( \frac{10^{-10} M_{\odot}}{M_0} \right)  \simeq
    2 \times 10^7 \mbox{s} \left( \frac{10^{22} \mbox{g}}{M_0} \right),
\end{equation}
resulting in values that are remarkably close to those adopted by, for example, \cite{GenST20} (see their Eq.~39). This close agreement, though reassuring, is rather coincidental, as our value results from a detailed treatment of Bondi accretion for stiff EOSs in full general relativity.  Regardless, results quite similar to (\ref{t_max4}) have been invoked by previous investigators (see, e.g., \cite{CapPT13,GenST20} and references therein) to constrain PBHs in the mass range $10^{-16} M_\odot \lesssim M \lesssim 10^{-10} M_\odot$ as dark-matter candidates.  

Corresponding to Eqs.~(\ref{tmax2}) and (\ref{t_max4}), the minimum accretion rate given by Eq.~(\ref{mdot_min}) can be evaluated to yield initially
\begin{eqnarray}
\dot M_{\rm min} & = & 2 \times 10^5 \frac{M_\odot}{\mbox{s}} \frac{\bar M_{\rm max}^2}{\kappa} \left( \frac{M_\odot}{M_{\rm max}} \right)^2 \left( \frac{M_0}{M_\odot} \right)^2  \nonumber \\
& \lesssim & 3 \times 10^{-9} \frac{M_\odot}{\mbox{yr}} \left( \frac{M_0}{10^{-10} M_\odot} \right)^2,
\end{eqnarray}
where, in the last step, we have again assumed (\ref{kappa_max}) and $M_{\rm max} = 2.3 M_{\odot}$.

To summarize, we provide estimates for the accretion times of black holes residing in neutron stars.  While our results are very similar to previously adopted values, they are based on a rigorous, relativistic treatment of spherical Bondi accretion for stiff EOSs. We also demonstrate that there exists a {\em maximum accretion time}, depending chiefly on the initial black hole mass alone, by which time the black hole will have consumed the entire neutron star. We further argue that actual accretion times will not differ much from the maximum accretion time, so that the latter provides an approximate universal value for the lifetime of a neutron star with a small black hole residing at its center.

While our discussion here centers on PBHs and their contribution to the dark matter, the same estimates apply to some alternative scenarios as well.  Specifically, the neutron star could also capture other dark matter particles which, under sufficiently favorable conditions, could form a high density object that collapses to form a small black hole in the neutron star interior and ultimately consume the entire star.  Invoking observations of neutron star populations, several authors have used these arguments to derive constraints and bounds on dark matter particles (see, e.g., \cite{GolN89,deLF10,BraL14,BraE15,BraLT18}).

Our arguments build on a number of assumptions, of course.  We assume that the accretion process is dominated by spherical, steady-state, adiabatic Bondi accretion (which may not hold for extremely stiff EOSs, e.g.~polytropes with $\Gamma > 3$; see, for example,~\cite{Beg78,RicBS21a}), and we ignore effects of rotation (which seems justified according to the findings of \cite{EasL19}; see also \cite{Mar95,KouT14}).  We also ignore radiation and radiation pressure, although the trapping radius for photons \cite{Ree78} is likely very large and radiation may be totally ineffective in holding back the hyper-Eddington accretion that can arise here (see, e.g., \cite{InaHO16}). Radiation processes, including the possible role of neutrinos, as well as the role of conduction, need to be probed further and the effects of fermion degeneracy must be taken into account. The possible effects of magnetic fields also have been ignored and should be investigated.  We implicitly assume that the weak dependence of the accretion rates and times on the polytropic index of the EOS indicate a similarly weak dependence on the detailed properties of realistic nuclear EOSs. There could, in principle, be phase transitions and other effects that might occur in the accretion flow, but we note that the Bondi critical radius occurs rather close to the black hole for typical cases (see Fig 2 in \cite{RicBS21a}). This means that the density and temperature in the accreting gas do not increase substantially above their core values before the gas is captured.  In spite of these considerations, we believe that our result provides an interesting limit on the timescale for the demise of a neutron star by an endoparasitic black hole, not least because it provides more rigorous justification for a key assumption used in constraining the contribution of PBHs to the dark matter content of the Universe.

\acknowledgments

It is a pleasure to thank Gordon Baym and Chloe Richards for interesting discussions.  This work was supported in part by National Science Foundation (NSF) grants No.~PHY-1707526 and No.~PHY-2010394 to Bowdoin College, and NSF grants No.~PHY-1662211 and No.~PHY-2006066 and National Aeronautics and Space Administration (NASA) grant  No.~80NSSC17K0070 to the University of Illinois at Urbana-Champaign.


%

\end{document}